\documentclass[12pt]{article}
\usepackage[T1]{fontenc}
\usepackage[utf8]{inputenc}
\usepackage{newtxtext,newtxmath}
\usepackage[margin=1in]{geometry}
\usepackage{setspace}
\usepackage{microtype}
\usepackage{array,booktabs,tabularx}
\usepackage[labelsep=period,labelfont=bf,font=small]{caption}
\usepackage{xurl}
\usepackage[colorlinks=true,linkcolor=blue,citecolor=blue,urlcolor=blue,hyperfootnotes=false]{hyperref}
\newcommand{\uploaddate}{November 20, 2026}

\title{Where the Model Frequently Meets the Road: Combining Statistical, Formal, and Case Study Methods\thanks{The authors thank Ariel Ahram, Macartan Humphreys, Alan Jacobs, and Jaye Seawright for helpful comments. The authors used an AI assistant (Claude, Anthropic) to help edit parts of this version of the manuscript.}}
\author{Andrew Bennett\thanks{Department of Government, Georgetown University; bennetta@georgetown.edu.}\\ Georgetown University
  \and Bear Braumoeller\thanks{Deceased May 3, 2023.}\\ The Ohio State University
  \and John Severini\thanks{Department of Government, Georgetown University; jcs386@georgetown.edu.}\\ Georgetown University}
\date{\uploaddate}

\begin{document}
\maketitle

\begin{abstract}
\noindent This paper briefly analyzes the working or default assumptions researchers in the formal, statistical, and case study traditions typically hold regarding the sources of unexplained variance, parameter values, functional forms, time, and external validity. We argue that these working assumptions are often not essential to each method, and that these assumptions can be relaxed in ways that allow multimethod work to proceed. We then analyze the comparative advantages of different combinations of formal, statistical, and case study methods, and offer advice on how to combine these methods.
\end{abstract}

\onehalfspacing

\section*{Achieving the Full Potential of Multimethod Research}

The great potential for synergistic research combining the three leading methodological traditions in political science---formal modeling, case studies, and statistics---is not being fully realized. There is much to be gained by combining alternative research methods, as each method can compensate for the limits of the others. Formal models can provide deductively valid and counterintuitive insights, but they are not an empirical method and must be combined with either statistical studies or case studies to become so. Case study methods can use process tracing to look for detailed evidence on causal mechanisms and causal processes within cases, but they face limitations in generalizing these results to cases not studied. Statistical studies can identify patterns and correlations in specified populations, but they lack the guarantee of logical theoretical consistency that formal models can provide, and they face the challenges of potential endogeneity, spuriousness, and ecological inference (i.e., even if a correlation does point to a causal relationship, that relationship does not necessarily explain any particular case in the population). Combining methods provides opportunities for the development and testing of theories that no single method can match.\footnote{For analyses of the contributions different methods have made to specific research programs, see Sambanis (\hyperlink{ref-JBAPDPC9}{2004}) on civil war, Bäck and Dumont (\hyperlink{ref-B5X6DKI6}{2007}) on political coalitions, Hafner-Burton and Ron (\hyperlink{ref-CDC9T5NE}{2009}) on human rights, and George and Bennett (\hyperlink{ref-6BUL5HFZ}{2005}) on the inter-democratic peace.}

Yet the promise of multimethod work has been only partly fulfilled because of methodological challenges, institutional barriers, and conceptual limitations. These include the technical difficulties of each method alone and the even greater difficulties of properly combining them. Researchers often simply use different methods in parallel as a way to double-check their conclusions through ``triangulation.'' But triangulation attempts to achieve the often-unattainable goal of confirming a single finding by independently addressing the exact same question multiple times. This parallel approach often leads to impasse if results diverge and if researchers lack a way to properly adjudicate conflicting answers. Moreover, the actual epistemic gains from triangulation are heavily contingent on the methods being employed in a reliably independent manner, a condition difficult to satisfy in observational social science, where errors and biases often occur in tandem (Kuorikoski and Marchionni \hyperlink{ref-VDZPQ7P8}{2016}, \hyperlink{ref-QIAY7TUF}{2023}, 1453).

Several recent frameworks have moved beyond loose triangulation toward genuine integration through a more unified causal framework. Seawright (\hyperlink{ref-54ZCW7YV}{2016}) proposes that methods be combined so that one method tests the assumptions underlying another's causal inferences, in turn addressing concerns about potential confounders, measurement validity, or causal pathway plausibility that a single method must otherwise assume away. Goertz (\hyperlink{ref-UA3ZA7YM}{2017}) argues that multimethod research can only work through causal mechanism analysis, with qualitative case studies, for instance, providing within-case evidence that more quantitative cross-case methods cannot generate. In a similar vein, Humphreys and Jacobs (\hyperlink{ref-6EJI8FVR}{2015}, \hyperlink{ref-TC45B3MR}{2023}) offer a formal Bayesian framework on how to update beliefs about causal effects by connecting within-case process tracing with cross-case correlational patterns. Fairfield and Charman (\hyperlink{ref-8288HE6Y}{2022}) develop a distinct Bayesian approach, providing concrete guidelines for case study inference and arguing that Bayesian reasoning can operate across research traditions.

Integrative multimethod frameworks thus share the premise that methods from different traditions contribute to a common inference when they address the same underlying questions using compatible assumptions about causality. Seawright (\hyperlink{ref-54ZCW7YV}{2016, 19--31}), for instance, argues that the now well-known potential outcomes framework provides such a shared standard. David Waldner (\hyperlink{ref-9I7Q33MH}{2026}) builds a bridge between case study methods and the potential outcomes approach by outlining criteria for case studies to achieve causal identification. In this approach, a case study can achieve valid causal inference by establishing causal continuity along the ``front-door path'' (an unbroken chain of mediators and mechanisms from the hypothesized cause X to the outcome Y of the case) and by ruling out ``back-door paths'' (pretreatment common causes of both X and Y) and ``side-door paths'' (causes of the outcome that are independent of X). Humphreys and Jacobs (\hyperlink{ref-6EJI8FVR}{2015, 654--60}) argue that Bayesian inference, combined with the potential outcomes framework and, in their later work, with causal models (\hyperlink{ref-TC45B3MR}{Humphreys and Jacobs 2023}), provides an intuitive shared standard for integrating process-tracing and correlational evidence because both provide information about the same underlying causal parameters.

But while properly integrated multimethod work framed upon a unified set of causal assumptions is worth aspiring to, it remains difficult to achieve in practice. Disciplinary incentives inhibit the co-authorships that multimethod work often requires and contribute to the relative neglect of some combinations of methods. Scholars using discourse analysis and those using ``text as data'' methods, for instance, share substantive interests but rarely collaborate because of their supposedly insurmountable epistemological differences (\hyperlink{ref-LU5M6EPK}{Bennett 2015}). There is also the danger that researchers carrying out multimethod work will be held to impossible standards, viewed as falling between three stools, not sufficiently expert in any one method to satisfy its specialists.

Perhaps the most pervasive inhibition on multimethod work, however, is conceptual: researchers in each methodological tradition have tended to share methodological default assumptions that clash with those adopted by practitioners of alternative methods. For a long time, these methodological, institutional, and conceptual barriers kept multimethod work from growing even as each methodological tradition became more refined. One survey found that from 1975 to 2000, the proportion of articles using more than one method remained steady, ranging between 15 and 19 percent of the articles surveyed. Moreover, these multimethod articles rarely used some combinations of methods. Most (between 50 percent and 90 percent in different time periods) combined formal models and statistical research, but only 33 to 50 percent combined statistics and case studies, and only 5 to 16 percent combined formal models and case studies (\hyperlink{ref-2BYY2P5T}{Bennett et al. 2003}). Multimethod research has grown considerably since then (\hyperlink{ref-54ZCW7YV}{Seawright 2016, 3--4}), which suggests that these barriers are weakening. Yet the most common designs still run case studies and statistical analyses in parallel. In one experiment, expert evaluators never rated multimethod articles significantly higher than at least one single-method version of the same article (\hyperlink{ref-QJM8YRXH}{Seawright and Koivu 2026, 145--46}).

We explore one explanation for this disjuncture: substantial and often implicit differences in the methodological assumptions that researchers from each tradition typically hold regarding the sources of unexplained variation and outlier cases, parameter values, functional forms, time, and external validity.\footnote{There is of course a diversity of opinion on these issues among methodologists and practitioners of each methodological tradition. We hope only to paint an accurate portrait of views commonly held within each tradition, not to imply that these views are universal within each research community. Also, many methods contribute to ``multimethod'' research, including experiments, surveys, ethnographic studies, simulations, and other approaches. We focus on the three general categories of qualitative case studies, formal models, and regression analysis for reasons of space.} Our central claim is that these differences reflect methodological working assumptions rather than epistemological or ontological commitments. We argue that these working assumptions are often not essential to each method, and that there is much to be gained by relaxing them in ways that allow multimethod work to proceed. We then analyze the comparative advantages of different combinations of methods for various theory-building and theory-testing research objectives. We illustrate these advantages and offer methodological advice on how to combine different methods.

\section*{The Clash of Assumptions}

A major obstacle to integration across methods is the seeming incompatibility among the assumptions widely held by practitioners of different methodological traditions. For instance, Mario Small notes that ``some authors have argued that combining quantitative and qualitative perspectives is not possible without contradiction, because different methods reflect different epistemologies, which, by definition, hold different assumptions about the nature of truth.'' Small adds that proponents of this view ``overstate how closely tied epistemological perspective and everyday practice are,'' but he then states that ``some analytical approaches are, in fact, incommensurable, because their techniques are tightly coupled with conflicting epistemological perspectives'' (\hyperlink{ref-3Z7UTEAC}{Small 2011, 77--78}).

Similarly, Amel Ahmed and Rudra Sil have cautioned that multimethod research risks incoherence when attempting to combine ``nomothetic'' approaches (i.e., formal theories or large-N studies seeking general laws) with ``idiographic'' ones (i.e., context-specific narratives). They argue that because these methods proceed from inherently different understandings of social reality, they can never truly cross-validate each other; as such, forcing them together may in fact undermine single-method scholarship (\hyperlink{ref-8R35HYVF}{Ahmed and Sil 2012}). Ariel Ahram (\hyperlink{ref-9DWGRC2H}{2013}) identifies a parallel measurement risk, warning that the ``thick,'' mechanism-laden concepts more often preferred by qualitative scholars can frequently be incompatible with the ``thin'' variables necessary for statistical analysis, in turn leading to ``mechanism muddling'' rather than valid synthesis. Derek Beach (\hyperlink{ref-BQK8DWJB}{2020}) reinforces this skepticism, arguing that the top-down logic of variance-based statistical inference may be effectively incommensurable with the bottom-up logic of case-based process tracing. Beach and Kaas (\hyperlink{ref-NX8M7JSM}{2020, 214}) go further, arguing that the differences among variance-based, case-based, and interpretivist methodologies are irreconcilable.

On this view, because variance-based and case-based approaches rely on fundamentally distinct assumptions about causality, frameworks attempting to merge them are more apt to produce category errors than achieve true integration. Perhaps most influentially, Goertz and Mahoney (\hyperlink{ref-SPJZYPYM}{2012}) characterize quantitative and qualitative research as constituting ``two cultures'' with distinct and internally coherent sets of norms, practices, and assumptions. While they do not claim outright incompatibility, their account emphasizes how deeply each tradition's operating logic differs from the other's.

In contrast, we argue that the assumptions dividing these traditions are typically convenient methodological defaults rather than epistemological or ontological commitments. Each tradition has a characteristic way of handling unexplained variance, parameter values, functional form, time, and external validity. These ways are only loosely tied to particular traditions, and researchers in any tradition can adopt them when their questions call for it.

None of these assumptions reflect deep commitments about the nature of social reality; rather, they are simplifications that may be relaxed as evidence warrants. Galvin and Seawright (\hyperlink{ref-7YXV537F}{2023, 1642}) take the same stance toward regression, treating an initial statistical model as error-laden and using case studies to discover and partially correct its errors. The same is true of the assumptions that ostensibly divide methodological traditions. When Beach (\hyperlink{ref-BQK8DWJB}{2020}) argues that variance-based and case-based inference employ fundamentally distinct logics, he identifies real differences in how these methods process evidence. But these are differences about~procedure, not about underlying assumptions of causality. As the above integrated frameworks demonstrate, both traditions can represent their central causal claims within a unified framework. For example, in the potential outcomes framework the causal effect of a treatment is defined as the difference between the outcome a unit experiences and the outcome it would have experienced otherwise. While such a definition is silent on how to estimate the effect, it provides a common language for researchers across traditions. Statistical methods might estimate average causal effects across populations; case studies might investigate whether and how a cause operated in a particular instance. These are different approaches to understanding the same potential outcomes; much of their apparent incommensurability dissolves once this is recognized.

In our view, most researchers share common epistemological assumptions rooted in scientific realism: a social world exists independent of our minds; our theories, explanations, and interpretations try to approximate how the causal mechanisms in that social world work to generate outcomes; and we assess the accuracy of alternative theoretical explanations by comparing how well they fit with outcomes and processes at the level of populations and individual cases. Recent guides to research design in political science like Blair et al. (\hyperlink{ref-8HEDZWYC}{2023}) operate squarely within this framework, as do most practitioners regardless of methodological training.\footnote{See also Chernoff (\hyperlink{ref-9U5Q86AS}{2007}).} The barriers to multimethod research arise squarely from treating default assumptions as if they reflected fundamental truths about what each method can or cannot do. It therefore behooves us to examine the key working assumptions held by researchers in each methodological tradition and to explore how they can be relaxed or adapted in ways conducive to multimethod research.\footnote{For a similar discussion of the differing assumptions of statistical and case study researchers, see Goertz and Mahoney (\hyperlink{ref-T3FGLHWP}{2006}).} We provide a roadmap for this in Table 1.

\begin{table}[htbp]
\centering\footnotesize
\renewcommand{\arraystretch}{1.3}
\begin{tabularx}{\textwidth}{@{}>{\raggedright\arraybackslash}p{0.12\textwidth}*{4}{>{\raggedright\arraybackslash}X}@{}}
\toprule
\textbf{Assumption} & \textbf{Statistical} & \textbf{Formal} & \textbf{Case Study} & \textbf{Resolution} \\
\midrule
Sources of variance & Unexplained variance treated as a practical given; focus on aggregate patterns & Mixed: pure-strategy models yield deterministic predictions; mixed-strategy and evolutionary models incorporate randomization & Outliers problematized; seek causal explanations for individual cases & Where to stop explaining variance is a pragmatic choice; multimethod designs help adjudicate \\ \addlinespace
Parameter values & Focus on population-average treatment effects; variables limited to preserve degrees of freedom & Simplifying assumptions on parameter magnitudes for tractability & Case-specific causal effects; tolerance for causal complexity and heterogeneity & Different estimands, not incompatible methods; complementary rather than contradictory \\ \addlinespace
Functional form & Driven by data type and statistical parsimony; increasing attention to nonparametric methods & Theory-derived but best practice seeks generality over specific parametric forms & Less explicit; process tracing can help adjudicate among functional forms & Case studies can check whether an assumed form matches the causal process in a case \\ \addlinespace
Time & Default implementations often treat timing as incidental; time-series and hazard models as exceptions & Time often incidental by default; evolutionary and dynamic models as exceptions & Sequencing, duration, and timing central to process tracing & Differing defaults about temporal complexity, not incompatible assumptions; each method addresses different temporal dimensions \\ \addlinespace
External validity & Traditionally presume representativeness via sampling; design-based inference increasingly focused on local estimates & Deductive generality assumed but scope conditions often underspecified & Generalization earned through case selection, scope conditions, typological theory & Complementary strategies: case studies probe scope conditions; statistics test generalizability \\ \addlinespace
\bottomrule
\end{tabularx}
\caption{Working Assumptions Across Methodological Traditions and Resolutions}
\label{tab:assumptions}
\end{table}

\subsection*{Assumptions about Sources of Unexplained Variance and the Significance of Outliers}

Whether couched as abstract arguments over the fundamental nature of causation or concrete arguments about the sources of error in the dependent variable, disagreements over assumptions about the sources and magnitude of error in the dependent variable are a persistent feature of methodological discourse.\footnote{As Hall (\hyperlink{ref-6YDIYBTN}{2003}) demonstrates, accounts of the empirical implications of causal relations dovetail all too well with descriptions of the empirical patterns that different methods are designed to unearth.} The most overt divergence here is between qualitative and quantitative methodologists. In much quantitative work, unexplained variation in the dependent variable is treated as given, and whether it reflects inherently stochastic processes or simply epistemic limitations is often pragmatically passed over. In addition, researchers often leave out of their statistical models many potential explanatory variables that they expect would have weak effects and whose inclusion would reduce the degrees of freedom and create other methodological problems. For this reason, the criticism that much unexplained variance remains, or that this or that case does not fit the story captured by the statistical model, is not seen as persuasive: it is to be expected that some cases will not fit the model, simply because their outcomes cannot be perfectly predicted. Even ``failure'' in a case in which the predicted probability of a ``success'' is 0.9 says little about the quality of the model, as such events should happen one time out of ten. Only the aggregate results are seen to be convincing.

At the other end of the spectrum, case study researchers focus to a much greater degree on deviant or outlier cases, or those whose outcomes do not fit the model. Rather than treating unexplained variation as a given, case study researchers problematize it. Absent some convincing rationale, they seek strong causal explanations of cases on the assumption that theoretically intelligible social behavior rather than chance accounts for outcomes.

Interestingly, formal modelers are of two minds on this issue. Many applied formal models in political science focus on pure-strategy equilibria that yield deterministic predictions, but observed social realities clearly do not neatly conform to this expectation. Thus, modelers have found it necessary to devise equilibrium concepts that can accommodate out-of-equilibrium observations. Consequently, one category of formal models, largely outside of the rational choice tradition, describes outcomes in stochastic terms. Agent-based evolutionary modeling is perhaps the best example of this approach (\hyperlink{ref-BMXZAIYA}{Weibull 1997}; \hyperlink{ref-KHSUM2S3}{Cederman 2001}). In this kind of model, the agents' ``success'' in their environment affects the frequency of reproduction, but the process is inherently stochastic. Randomization is also central to rational choice theory, since Nash's existence theorem relies on mixed strategies. A basic mixed-strategy equilibrium, for example, is based on the idea that each player will randomize in order to keep the other indifferent between one strategy and the other.

Even in cases in which the model points to a dominant strategy for all players, formal theorists have posited that some randomization may occur as the result of variations in utility that are observed by the player but not by the scholar.\footnote{See Smith (\hyperlink{ref-Y7NHXXZ5}{1999}) for a clear discussion of this point.} Due to such variations, players are assumed to choose those strategies deemed optimal by theorists with a greater probability than they would non-optimal ones, but not with a probability of 1. This is the concept of a quantal response equilibrium (QRE) (\hyperlink{ref-55UHRDCB}{Morton 1999}).

Despite these apparent divergences, differences in assumptions about probabilism and determinism need not impede multimethod research. As King, Keohane, and Verba (\hyperlink{ref-UFX8R23S}{1994, 59--60}) argue, we cannot ultimately know whether the social world is deterministic, with unexplained variation reflecting omitted causes, or inherently stochastic. The differences across traditions are not as sharp as they appear, since many statisticians may in fact hold closer to a deterministic perspective in which unexplained variance could in principle be accounted for by omitted variables.

A researcher can in fact pursue statistical research with deterministic assumptions, or case study methods with probabilistic ones. Charles Ragin's ``fuzzy set'' methods, for example, can be interpreted as a way to allow probabilistic measures in qualitative work (\hyperlink{ref-TQGN6566}{Ragin 2000}). What really matters is not any general assumption a researcher makes about whether the social world is stochastic, but the pragmatic implicit working assumption that the researcher makes when he or she decides at what point to stop in seeking to reduce or explain the error term in a particular research project. If a researcher using statistical methods decides to explore whether adding one more variable will improve their model, they are implicitly assuming that the world may be at least a little more deterministic than their initial model indicates. If a researcher undertaking a case study decides to stop pursuing every anomalous case or every unexpected aspect of a case, they are behaving no differently than if they had decided that this is the place to allow for the possibility that the remainder of the error term is generated by a stochastic process.

The real question, then, is how to decide when to stop pursuing an explanation of the error term, and it is on precisely this question that the advantages of multimethod work are evident. For case study researchers, a common working assumption is that outliers are fruitful cases to study for the purpose of inductively finding left-out variables, overlooked interaction effects, or mistaken assumptions about the functional form of the relationships among the variables. Case study researchers are also attentive to the possibility that measurement error may lead a case to be misclassified as an outlier, but they tend to downplay the possibility that stochastic processes may also account for outliers. It may also turn out, however, that large outliers represent nothing more than a case with a very substantial error component---a perfectly normal, even expected, occurrence. By the same token, a dramatic misprediction plus random noise could just as easily produce an artificially small outlier. So rather than making an assumption about the (in)significance of cases with high error terms, the researcher can combine methods to assess outliers.

A researcher using statistical tests who suspects that the error term may be legitimately reduced by adding an additional variable might look at an outlier case (or several such cases) to uncover a candidate variable, or justify an added variable by demonstrating that it not only improves the statistical fit but is consistent with the process-tracing evidence from intensive study of a case from the sample. A case study researcher worried that s/he might be generating a ``just so'' story in trying to explain an anomalous case by adding a variable might use a statistical test to assess whether the additional variance explained outweighs the decrease in parsimony, or more ambitiously, whether the new variable improves the fit of the model. A game theorist who hypothesizes that one branch of a game tree resembles a Prisoners' Dilemma but who finds through statistical analysis that the variance of outcomes at that branch is significantly greater than anticipated might reasonably suspect either that the measured payoffs do an unusually poor job of capturing actual payoffs or that the structure of the game at that branch has been misspecified. A detailed heuristic case study or two could go a long way toward improving understanding of the players' behavior at this juncture.

Here again, we argue that researchers from each tradition should focus on the more fundamental possibility that in principle the outliers in any particular piece of research could be any of the above: the product of stochastic processes, the result of measurement error, the consequence of a variable that affects the outcome of the outlier case and also those of other cases in the sample, or the consequence of a variable that is unique to the outlier case or nearly so and that may have explanatory power for the case even if it does not significantly improve the statistical fit for the sample.\footnote{Another way to put the differences among researchers regarding outliers concerns their different views of the ``unit homogeneity'' assumption. Statistical researchers and formal modelers tend to assume that unit homogeneity, or at least the weaker version of constant causal effects, will hold for large populations of cases. Case study researchers are more sensitive to the possibilities of ``equifinality'' (multiple causal paths to the same outcome), ``multifinality'' (multiple outcomes associated with any one value of a variable), and general heterogeneity among cases that are influenced by numerous variables and interactions among these variables. Thus, case study researchers are more likely to break populations down into narrower subtypes within which unit homogeneity is more likely to hold, such as types of democracies (or ``democracy with adjectives'' (\hyperlink{ref-LFPCNBTR}{Collier and Levitsky 1997})).} Combining different methods provides more powerful means than any one method used alone for discerning which of these holds true. Statistical methods help identify which cases are outliers relative to a particular model. Case studies of outliers can help test if measurement error is involved and can also identify candidate variables that might explain why the outcomes in these cases deviate from the model's expectations. Formalization of these variables can help elucidate their logic and clarify the forms they might take in statistical equations. Statistical studies using these new variables (and omitting the cases from which they were derived) can then help determine if they improve the fit for a wider sample of cases. Even if a variable inductively identified from a case does not improve the statistical fit for the rest of the sample, it might still explain the outlier case if it is relatively unique to that case and if it is backed up by convincing logic and process-tracing evidence from that case (as when a voter's unexpected choice is explained by a personal connection to the candidate that is otherwise idiosyncratic but causally real).

A more difficult methodological challenge is that it is hard to identify cases that are in fact outliers relative to a statistical model but whose deviation is masked by measurement error or random noise. It might be possible to identify such cases by exploring whether statistical results are robust in the face of slightly different measurements of the variables. Even if such different measures have roughly equivalent fit for the whole sample, they may point to different cases as possible outliers for examination in detailed case studies.

\subsection*{Parameter-Value Assumptions}

Researchers from different traditions often target different causal quantities without recognizing that they are doing so, leading to apparent contradictions that are in fact complementary findings about different aspects of the same phenomenon. Anyone who has engaged in formal modeling realizes that model implications become dramatically more complex as included variables increase. Moreover, readers can lose interest when the implications of the model can only be expressed in page-long systems of convoluted equations. The temptation exists, therefore, to make life (and equilibrium solutions) easier by making a few simplifying assumptions about the magnitude of some of the model's parameters, either in absolute terms or relative to one another.

Statistical researchers are more likely to estimate the average treatment effect of X on Y across a population, while case study researchers are more likely to examine whether X caused Y in a specific case. These analyses can often yield apparently contradictory results: a statistical estimate could be significant and positive while a case study finds no evidence that X operated causally in a particular case. Parameter heterogeneity is often one cause of this disagreement; the causal effect of X could vary across cases and the particular case studied may happen to have a small or null effect or even an opposite effect (as when one person is allergic to a medicine that is helpful for most people). Indeed, both findings can be simultaneously correct because they target different estimands: the population average treatment effect versus the case-specific causal effect. A unified causal framework like potential outcomes makes this obvious by explicitly defining case-level effects $Y_{i,t} - Y_{i,c}$ as potentially varying across units (\hyperlink{ref-W87IX4B8}{Holland 1986}; \hyperlink{ref-8E9EIDTF}{Rubin 1974}), but these working assumptions remain an obstacle to integration because they often result in counterintuitive and ostensibly contradictory findings.

Differences between approaches to degrees of freedom reflect practical differences in how each tradition conceptualizes the problem of inference. Statistical researchers often worry about overfitting and try to limit the number of independent variables in their models to preserve degrees of freedom. In contrast, case studies that analyze evidence from a case with Bayesian rather than frequentist methods face no degrees-of-freedom problem when alternative explanations add independent variables. Complexity still carries a cost, because more complex explanations start with lower prior probabilities and must overcome that penalty with stronger evidence (\hyperlink{ref-8288HE6Y}{Fairfield and Charman 2022, 239--43}). Adding independent variables in case study analyses does add combinatorial complexity to the types or configurations of cases that are possible. Case study researchers can use typological theorizing, however, to both logically and empirically narrow the focus to the combinations of variables, or types, that provide strong comparisons, interesting outliers, or other kinds of cases worthy of in-depth analysis (\hyperlink{ref-6BUL5HFZ}{George and Bennett 2005}; \hyperlink{ref-T9W6J9NE}{Elman 2005}).

Parameter-value assumptions also matter for interaction effects. As Goertz (\hyperlink{ref-UA3ZA7YM}{2017, 24, 123}) notes, interactions are claims about causal mechanisms, that is to say, a claim that $X_1$ and $X_2$ work together to produce Y. On the statistical side, interaction terms can create interpretive difficulties. Interaction models are often misunderstood because the coefficients on $X_1$ or $X_2$ are looked at in isolation even though the substantively important quantity is the marginal effect of one variable conditional on values of the other variables. Researchers risk misinterpretation when they fail to report marginal effects over the observed range of the conditioning variable (\hyperlink{ref-MEVHZHP6}{Brambor et al. 2006}). Case studies investigating hypotheses involving interactions must examine both components and their joint operation, yet standard advice on case selection for multimethod research rarely addresses interaction-term hypotheses, instead treating independent variables as univariate. When leverage comes from comparisons across cases, interaction hypotheses ideally require cases spanning all four quadrants of the $X_1 \times X_2$ space: cases where both are high, both are low, and each combination of high and low. Selecting only cases where both $X_1$ and $X_2$ are present (the upper-right quadrant) cannot distinguish an interaction effect from additive main effects. More explicit attention to interactive causal mechanisms in case study design would strengthen the integration of formal and qualitative methods.

The fundamental point for all researchers is that while theoretical parsimony is desirable, our theories need to be as complex as social reality demands (\hyperlink{ref-UFX8R23S}{King et al. 1994}). In a maxim often attributed to Einstein, a scientific theory should be ``as simple as possible, but no simpler.'' If a social phenomenon appears to require more explanatory variables than we can wrap our minds around, it is often better to define subtypes of the phenomenon more narrowly, control for some of the variables, and theorize around more narrowly defined types, rather than omit potentially relevant variables or engage in conceptual stretching. In deciding where to partition a phenomenon into relevant subtypes, empirically observed statistical distributions, logic, and case studies of the ideal type and boundary cases all have roles to play.

\subsection*{Functional Form Assumptions}

We use the term ``functional form'' broadly, to include both the shape of the relationship between causes and the expected outcome and the distribution of outcomes around it. The strength of functional form assumptions varies widely across methodological techniques. Case study researchers tend to be less explicit and precise about the functional forms their theories lead them to expect to find. When a researcher is dealing with many observations within a given case---as in content analysis, descriptions of interview results, etc.---the tendency is to ``let the data speak'' by presenting graphical trends or marginals and discussing the implications for theory. When researchers are explicit about the functional forms of the relationships they expect to find, however, it may be possible to winnow these forms down on the basis of process-tracing evidence. Even when many alternative functional forms are equally consistent with the outcomes observed in the cases, they may not be equally consistent with the observable processes through which these outcomes arose.

Formal theorists, on the other hand, generally hold that the functional form of an empirical test should follow from the deductive theory motivating it. Yet good formal theorists also seek generality and try to minimize dependence on specific functional forms, preferring (for instance) general cost functions characterized by their qualitative properties over specific parametric forms. The tension between these impulses means that in practice, formal models vary considerably in how tightly they constrain functional form.

Users of statistics fall between these two relatively defensible stools. Traditionally, functional form assumptions have mostly been based on the most appropriate statistical technique given the nature of the data. If the dependent variable is binary, a logit or probit functional form is chosen; if a count variable, then Poisson; if continuous and unbounded, some flavor of (typically linear) regression. These techniques were, typically, originally designed with statistical parsimony rather than theoretical generality in mind. The implications of that fact are, one, that their mathematical properties are typically well understood, and two, that they may or may not have anything to do with the functional form implied by the theory actually being tested.

Attempts to rectify this disjuncture between assumptions and theory, on the one hand, and data, on the other, have led large-N methodologists in two opposite directions. The first involves tailoring functional form assumptions more carefully to the implications of theory: Signorino (\hyperlink{ref-TMDYM4QZ}{1999}) and Smith (\hyperlink{ref-Y7NHXXZ5}{1999}) have sought to capture the empirical implications of game-theoretic models, for example, while Braumoeller (\hyperlink{ref-WQHW3M49}{2003}) has attempted to give statistical meaning to the case study researcher's argument that multiple causal ``paths'' can lead to the same outcome. The main point here is that functional form assumptions carry implications for causal identification. A regression assuming linearity when the true relationship is in fact nonlinear will yield biased estimates of causal effects, potentially even reversing the sign of the true effect in some parts of the data. As Morgan and Winship (\hyperlink{ref-4LFCHVCD}{2014, chaps. 5, 6}) emphasize, regression and matching attempt to achieve causal identification through conditioning on observed covariates, with success depending on correct model specification. In parametric models, assumptions about confounding thus interact with functional form in complex ways. A confounding variable that would bias a linear model might not bias a correctly specified nonlinear model, and vice versa. Nonparametric approaches to confounding, such as conditioning on all levels of a confounder, can avoid this dependence. For multimethod research, case studies investigating outliers or influential cases should care not only about omitted variables but also about whether the functional form assumed in the model matches the causal process actually operating in the case.

The alternative is to relax functional form assumptions as much as possible. Maximum-likelihood techniques, for instance, make distributional assumptions (probit assumes a cumulative normal error density) that, if wrong, can produce inconsistent estimates. Solutions range from estimating additional shape parameters (e.g., ``scobit''), to quasi-likelihood methods that dispense with distributional assumptions entirely, to Bayesian techniques that estimate posterior densities directly. Each adds flexibility at the cost of additional data requirements.

Perhaps the most extreme example of assumption-free modeling is matching, in which a variable measuring the individual's propensity for having been exposed to the ``treatment'' is used to create two matched samples of individuals, one of which was ``treated'' and the other not (\hyperlink{ref-LRF4UJB6}{Rubin 1973}; \hyperlink{ref-VPMVSA8W}{Rosenbaum and Rubin 1983}). Once the samples have been created, the effect of treatment can be measured via a simple comparison of means. The benefits of such an approach are substantial (especially since the robustness tests that can be applied are remarkably sophisticated). Even this most assumption-free technique must rely on some assumptions to generate its results: the stable unit treatment value assumption (SUTVA), which requires that the outcome for individual i not be altered by the treatment status of individual j, is one, as is ignorability, the assumption that no omitted covariate affects both treatment assignment and the outcome.

Regardless of the radical disparities between these two schools, they do agree up to a point on the question of functional-form assumptions: both would argue that too many assumptions are made for the sake of mathematical convenience rather than theoretical meaning, and that the inferences derived from such models are not to be trusted. Some, like normality of errors, are typically made for no theoretical reason and are easily violated. The bone of contention between the two groups lies in what to do with theoretically derived functional-form assumptions: one group prefers to model them as carefully as possible, while the other would eliminate them.

SUTVA violations are particularly prevalent in political science applications and pose a pragmatic problem for statistical inference that case studies can help address. Morgan and Winship (\hyperlink{ref-4LFCHVCD}{2014, 50}) describe two interrelated features of typical violations: influence across units through social or physical contact, and dilution or concentration of effects as treatment becomes more or less prevalent. For example, consider a job training program. In a large labor market, wages offered to retrained workers may be unaffected by the program's existence, but in a small labor market, wages may respond as trainees move in and out of the program (50). This would violate SUTVA. Political phenomena involving strategic interaction, policy diffusion, or general equilibrium effects face analogous challenges.

Case studies can be useful for working out assumptions latent in formal models or derived from statistical findings. If SUTVA is violated, there should be detectable evidence of spillovers in close case examination---actors responding to others' treatment status, strategic anticipation of others' choices, or market-level adjustments to treatment prevalence. If ignorability fails, confounders operating in case-level causal processes may be detectable. Case studies embedded in matching designs can thus test assumptions that statistical diagnostics alone cannot address (\hyperlink{ref-54ZCW7YV}{Seawright 2016, 111--17}).

\subsection*{Assumptions about Time}

Standard implementations of formal and statistical models often treat time as incidental rather than constitutive. While both traditions have developed sophisticated tools for temporal dynamics, their default applications tend toward what Kreuzer (\hyperlink{ref-ZRJSESUW}{2023, 116--17}) calls a ``linear'' conceptualization of causality, in which causal effects are assumed to be immediate, symmetric, and independent of temporal context.

Dynamic formal models and their statistical counterpart, time-series analysis, are designed to model situations in which a cause in one time period produces an effect in another, typically the next. The time period in question can be a day, a month, a quarter, a year, or whatever; time in the usual sense is typically not as important as ensuring that a relationship among variables that plays out over time is modeled appropriately. At other times, however, time itself is the phenomenon of greatest interest, as in evolutionary game theory or theories that examine the durations of wars or crises using duration or hazard models. In the former case, time is an important independent variable because it permits optimal strategies to be rewarded and suboptimal strategies to be punished, and it permits the reproductive repercussions of those rewards and punishments to unfold. In duration or hazard models, time is an important dependent variable in its own right.

For case study researchers, sequencing and timing are critical sources of inference through the method of process tracing. Even if the outcomes of different cases are similar, and are equally consistent with competing explanations, these outcomes may have arisen through different sequential processes, and the differences may be relevant to the explanation of the case as well as to its future trajectory. Case study researchers are thus attentive to the possibility of path dependency, and frequently look for indications that self-reinforcing dynamics or self-denying prophecies are at work (\hyperlink{ref-RWVRXHLP}{Pierson 2004}; \hyperlink{ref-U6ZZ5C9U}{Mahoney 2000}; \hyperlink{ref-8EDRRZ96}{Bennett and Elman 2006}). Indeed, the vast majority of work using theories of path dependency uses case study methods rather than quantitative methods.

Grzymala-Busse (\hyperlink{ref-GGQP5X7F}{2011}) provides one method-agnostic framework for distinguishing aspects of temporality relevant to causal inference.~\emph{Duration}~refers to how long an event or process takes. Some causal mechanisms require extended time to operate; rapid institution-building, for instance, is unlikely to involve deliberation or consensus among wide constituencies (1271).~\emph{Tempo}~refers to the rate of change per unit time, while \emph{acceleration}~refers to changes in tempo over time. The rate at which change unfolds constrains which actors can respond and often structures how those actors respond in the first place.~\emph{Timing}~refers to when an event occurs relative to other events or to a reference point. The availability and costs of strategies depend on when action is taken: early movers face different constraints than late adopters.

Causally focused distinctions like these clarify the relative strengths and weaknesses of different methods and the potential hurdles to integration. A case study investigating a statistical outlier should not only care about whether the hypothesized cause was present but also about its temporal structure by focusing on aspects of the case like how quickly it unfolded, when it occurred relative to other relevant events, and whether its effects accumulated gradually or happened all at once. Statistical models that ignore these temporal dimensions might otherwise properly identify population-level correlations while simultaneously missing case-level mechanisms that depend on duration, tempo, or timing. The deeper issue at hand is not necessarily that formal and statistical methods are incapable of properly handling time, but that their standard implementations often fail to do so. When researchers estimate cross-sectional models, they implicitly assume that treatment exposure timing is irrelevant, that order is unimportant, and that causes are operating instantaneously. These are strong assumptions that may be appropriate in some contexts but are not universal. Relaxing them is a methodological choice. Statistical time-series and hazard models can identify temporal patterns in populations. Formal models can clarify the logical structure of path-dependent or timing-sensitive processes. Case studies can provide within-case evidence about hypothesized temporal mechanisms like increasing returns, critical junctures, and sequencing effects---indeed, almost all of the empirical research that theorizes such path-dependency mechanisms uses case studies. Iteration across methods may help further refine inference about temporal mechanisms. Statistical analysis can reveal how outcomes differ for early versus late adopters, case studies the mechanisms responsible for this difference, and formal models the conditions under which certain advantages obtain.

\subsection*{External Validity Assumptions}

One substantial but often implicit clash of assumptions between case study and statistical research concerns external validity, or the conditions under which findings can meaningfully generalize beyond the cases examined in a study. For Findley, Kikuta, and Denly (\hyperlink{ref-I8J294LG}{2021}), research design often involves inherent trade-offs between external and internal validity across a variety of dimensions. While they focus chiefly on the inferential gap between experimental and observational studies, there are also key differences between how specific methods approach external validity concerns. Formal theory typically abstracts from empirical contexts to identify general mechanisms and specify ex ante which dimensions should be held constant or varied; statistical analyses optimize breadth across units and treatments but often leave mechanisms and timing implicit; and case studies excel in demonstrating mechanistic and temporal validity through purposive selection of theoretically informative cases.

Case study methodologists are acutely aware of the problem of external validity and generally skeptical of assumptions that non-trivial single-variable causal processes apply in unadulterated forms to large populations; in their world, credit for the ability to generalize beyond the cases studied must be earned. Typically this is done in one of three ways. First, a better understanding of a hypothesized causal process in a particular case, often developed partly inductively through close study of the case, may itself provide clues on the scope conditions of the process, or the types of cases, populations, or contexts to which this process is relevant. If we find, for example, that an individual voted for a candidate despite differences in party and ideology because the candidate was a blood relative, this provides an explanation that should only generalize to candidates and voters who are blood relatives (or perhaps more diffusely, neighbors, personal friends, and co-workers). Conversely, an improved understanding of a process developed through close study of a case may be very broadly generalizable. Charles Darwin's study of a small number of species in the Galápagos Islands, for example, helped him develop a theory that by its very nature should apply to all living things. For any explanation derived in part inductively through the study of a case, it is impossible to make an informed assessment of the generalizability or scope conditions of that explanation prior to developing the explanation itself.

A second and more deductive approach is to use extant theories to carefully specify a type of case that represents a particular combination of variables within certain ranges of values, and to then study that type of case with the presumption that any findings will take the form of ``contingent generalizations'' that will apply mostly or only to cases of this same type. Bennett and George discuss this approach as the development of ``typological theories''; Charles Ragin has talked about similar methods of treating cases as ``configurations'' of variables, and David Collier and Steven Levitsky note the development of a literature on ``democracy with adjectives'' in which each additional adjective denotes a subtype of the overall concept of democracy (such as a ``parliamentary'' versus a ``presidential'' democracy) (\hyperlink{ref-6BUL5HFZ}{George and Bennett 2005}; \hyperlink{ref-NDCA5MPS}{Ragin 1987}; \hyperlink{ref-LFPCNBTR}{Collier and Levitsky 1997}). In these approaches, the goal is to develop more internally valid and intensive concepts and hypotheses even if this comes at the expense of being able to apply them only to narrower types of cases.

A third approach to generalizing from cases is either to identify cases that disprove claims of necessity or sufficiency or to search out instances in which a theory fails to explain a case in which it is ``most likely'' to apply or succeeds in explaining a case in which it was ``least likely'' to hold true. If any such cases exist with respect to the phenomenon under study, even a single case can provide presumptive evidence that a claim of necessity or sufficiency is unfounded, or that a theory that fails in a most likely case should have narrower scope conditions and one that succeeds in a least likely case probably deserves wider scope conditions. An example of the latter is research in which alliance dependence was the only explanation to correctly predict the behavior of Germany and Japan in the 1991 Gulf War (\hyperlink{ref-XPSIAPBR}{Bennett et al. 1994}), which suggests that the alliance dependence variable might be a powerful one in any cases in which it is present.\footnote{Strictly speaking, the 1991 Gulf War was not a completely least likely case for the alliance dependence hypothesis. Even though all other hypotheses pushed against the actual outcomes in Germany and Japan, these countries were highly dependent on the United States for their security. Had Germany and Japan been only modestly dependent on the United States, and had it still succeeded in explaining their behavior despite the effects of all the other variables of interest, we could infer that the alliance dependence variable was even stronger and more generalizable.} In all three of the approaches to generalization from case studies, however, any findings on scope conditions are suggestive rather than definitive, as they could be undermined by measurement error or specification error in the cases studied. This is why studies of additional cases, whether through case study or statistical analysis, are valuable means of verifying the scope conditions of a theory that emerge from the analysis of case studies.

Researchers utilizing statistical methods, on the other hand, have traditionally tended to view their set of cases as a sample representative of a larger universe of cases.\footnote{Very rarely, though, is this universe of cases explicitly defined, and as a result the relationships are treated as though they were in fact universal. Not many researchers would actually argue that their universe of cases expands back to the beginning of time; most would have some qualms about generalizing back to 1648, the traditional date given for the debut of Westphalian sovereignty.} For that reason, claims of external validity typically rest on the sample, whether a random sample or as much of the population as the data allow. To take an example, Reiter and Stam's findings on democratic efficacy in warfighting are based mostly on the behavior of a very small number of democracies (\hyperlink{ref-X6G2SIW5}{Reiter and Stam 2002}). Being conscientious researchers, the authors included every control variable that they could think of that would make these countries unique or unrepresentative of democracies in general. Having done so, they were comfortable making generalizations about the larger population of democracies as a whole.

Researchers working in the case study tradition might question the assumption of unit homogeneity in this research and object that Reiter and Stam's conclusions do not extend much beyond these few democracies, regardless of which controls have been entered.\footnote{In fact, they did so, at a presentation at Harvard's Olin Institute.} Whether one believes the authors or their qualitative critics depends in large part on how far one is willing to extend the benefit of the doubt on external validity, and why. One means for judging such debates is to apply the case study standard noted above and ask whether there is some theoretical reason to believe that Reiter and Stam's cases were ``least likely'' cases, or tough test cases, or most likely/easy test cases for their theory. Another approach would be to address the implicit unit-homogeneity argument head-on: for the critique to hold, the relationship between democracy and victory would have to be different in the countries that dominate the Reiter--Stam study from that in other countries. The authors' control variables bear on this possibility only to the extent that they capture how the relationship between democracy and victory differs across countries. Similarly, it is incumbent upon critics not just to argue that other democracies are different but to specify how they differ, so that the importance of those differences can be investigated. A third technique, which can be quite convincing, involves estimating the model on one subset of cases and then assessing its external validity by testing its predictions in another, different subset. Braumoeller (\hyperlink{ref-2TZR38YI}{2008, 86--87}) provides an example of this approach. Finally, in any case in which unit homogeneity is at issue, straightforward equality-of-coefficient tests can typically help determine the validity of the assertion that the relevant coefficients differ across subsets.

\section*{Comparative Advantages and Limits of Combinations of Methods}

If the default assumptions of different methods can be relaxed in ways that make them compatible, as we have argued, what are the most fruitful ways to combine different methods, and the advantages and uses of different combinations of methods? We now turn to these questions, building upon the typologies of different kinds of theory-building developed by Arend Lijphart and Harry Eckstein (\hyperlink{ref-FUQJL7CL}{Lijphart 1971}; \hyperlink{ref-Y4Q2EFLU}{Eckstein 1975}). Although their typologies and terms are not exactly the same, Lijphart and Eckstein identify four theory-building goals or types of study that are useful for present purposes: plausibility probes, disciplined-configurative studies, heuristic or theory-generating studies, and theory-testing studies (this sets aside purely descriptive or idiographic studies, as statistical and case studies can be descriptive but neither would be used with a formal model for descriptive purposes only). Plausibility probes are ``quick and dirty'' studies to determine whether more laborious or expensive studies are justified, and to make changes to the research design before investing greater effort. Disciplined-configurative studies use a theory to explain a case or a statistical correlation whose variables or outcomes are already well known but whose result may have heretofore remained puzzling. Heuristic studies use deductive and/or inductive logic to identify new or left-out variables or interactions. Theory-testing studies test whether theories apply to or explain particular cases or populations of cases (strictly speaking, they test whether theories fail to apply, or at least fail worse than the available alternative theories).

Different sequences of methods are suited to different purposes. Starting with a formal model and then moving to statistical analysis enables theory testing, while starting with statistical analysis and then developing a formal model based on the findings fits the heuristic purpose of building new models. Evan Lieberman's analysis of multimethod research focuses on starting with a statistical analysis and using this to help select case studies for process tracing (\hyperlink{ref-VCIMU9U8}{Lieberman 2005}; \hyperlink{ref-GHM9DK79}{Gerring and Seawright 2006}). Too often, studies that rely on statistical analysis alone neglect ``outlier'' and ``influential'' cases. Outlier cases, or cases with a high error term in a statistical analysis, can be a fruitful area for doing case studies to look for new theories and potential omitted variables. Influential cases, or cases whose removal from a statistical analysis would have the greatest effect on the parameter estimates of the statistical model, can sometimes have large effects on parameter estimates. These are good candidates for case studies to assess whether the hypothesized mechanisms of the statistical model are evident in the case---if not, and if further investigation reveals that the influential case differs from the rest of the population in important respects, removing the influential case from the statistical analysis is a legitimate way to improve the original statistical theory's fit with the rest of the population.\footnote{Outlier cases may or may not be influential cases, and influential cases may or may not be outliers.}

Starting with statistical analysis and then moving to case studies, as Lieberman outlines, is a useful and common sequence, but case studies can also usefully precede statistical analysis. Case studies can develop theories inductively and clarify the attributes that constitute the relevant population before investing the time and resources necessary for coding numerous cases for statistical study (\hyperlink{ref-Y4Q2EFLU}{Eckstein 1975}; \hyperlink{ref-BD7XLSSF}{Rohlfing 2008, 1497}). It is also possible to do more than one iteration among methods when they are used in combination, and such work can serve more than one theory-building purpose in a single study. For example, one can generate a theory from a case study, test it more generally in a statistical population, and then do case studies on the outliers to further develop the theory. A formal model might be deductively derived and then tested against statistical data, which might lead to modifications to the model, which in turn can lead to tests against different or newly coded statistical data. Such iterations provide legitimate, progressive theorizing rather than illegitimate ``curve fitting'' as long as researchers are careful not to move from one statistical test to another in ways that create a lack of independence of tests. Iteration in successive case studies in a sequence of different methods largely avoids this problem, as process tracing in case studies works through a Bayesian logic, and the analysis of case study evidence is the same whether the underlying theory was developed deductively or drawn inductively from the case (\hyperlink{ref-8288HE6Y}{Fairfield and Charman 2022, 465--75}).

Building on the Lijphart--Eckstein typology, and keeping the issue of sequencing in mind, Table 2 summarizes the comparative advantages and uses of different combinations of methods. We discuss each combination in turn.

\begin{table}[htbp]
\centering\footnotesize
\renewcommand{\arraystretch}{1.3}
\begin{tabularx}{\textwidth}{@{}>{\raggedright\arraybackslash}p{0.16\textwidth}*{3}{>{\raggedright\arraybackslash}X}@{}}
\toprule
 & \multicolumn{3}{c}{\textbf{Type of Study}} \\
\cmidrule(l){2-4}
\textbf{Theory-Building Goal} & \textbf{Formal–Statistical} & \textbf{Formal–Case Study} & \textbf{Case Study–Statistical} \\
\midrule
Plausibility Probe & (A) See if re-doing models and/or data sets is worth the effort & (B) Identify changes to model or choose among models before further testing & (C) Identify changes to the model or which cases to study next \\ \addlinespace
Disciplined Configurative & (D) Use correlations to guide construction of a model, or model to explain a correlational finding & (E) Use a model to understand a case or a case to modify a model & (F) Use statistical findings to structure study of a case, or a case to identify variables for statistical models \\ \addlinespace
Heuristic / Theory-Generating & (G) Use a formal model to identify new or counterintuitive variables for statistical study & (H) Use deduction (model) or induction (case) to identify new variables to use in the complementary method & (I) Use statistical findings to identify outliers, then study these cases to find new or left-out variables. \\ \addlinespace
Theory Testing & (J) Statistical tests of formal models & (K) Case study / process tracing tests of formal models & (L) Cases test if putative processes behind correlations took place, statistical study tests if case findings apply to a population \\ \addlinespace
\bottomrule
\end{tabularx}
\caption{Comparative advantages of combinations of methods}
\label{tab:combinations}
\end{table}

\subsection*{Combining Formal and Statistical Methods}

The formal-statistical pairing offers several distinctive advantages. Formal models generate precise, testable predictions; statistical tests assess whether these predictions may hold across many cases, in turn providing evidence of generalizability that formal--case study combinations cannot easily achieve. Comparative statics derived from equilibrium analysis can also quite naturally translate into hypotheses about covariation. However, this combination also faces some key limitations. The sorts of mechanisms and assumptions central to formal models (i.e., beliefs, payoffs, information structures, discount rates, etc.) can be quite difficult to measure in large-N data, requiring researchers to rely on distant proxies (\hyperlink{ref-UA3ZA7YM}{Goertz 2017, 165--66}). Multiple causal processes may also generate similar observable patterns; as a result, statistical confirmation of a model's predictions does not necessarily establish the actual operation of the theorized mechanism (\hyperlink{ref-PQ5PGUWT}{Clarke and Primo 2012, 118--22}). Formal-statistical work often serves a heuristic purpose in which a formal model is used to derive deductively a new variable or specification, which is then tested statistically. These studies, while generally exemplary in their methodologies, miss opportunities for further developing their theories with little additional effort by paying attention to the outliers in their statistical tests.

More generally, researchers using statistical methods and formal models typically have very different views of ``outliers,'' and both groups could benefit from considering the views of case study researchers on this subject. Statistical researchers are prone to looking at outlying cases as the result of random errors, measurement error, or unique and identifiable causes, so that these cases can be safely ignored or should even be excluded from statistical analysis. Formal modelers whose models yield deterministic predictions tend to view outliers as impossible cases that, if they in fact occur, indicate a flaw in the formal model. Case study researchers would place more emphasis on the fact that formal and statistical models are always incomplete, so that outlier cases are potentially very fruitful grounds for identifying left-out variables and specification errors (or alternatively confirming that measurement error is indeed the source of the outlying observations, or finding that no suitable explanation exists, which may indicate either randomness or the limits of current knowledge). Researchers who have combined statistical methods and case studies have sometimes been very conscientious in identifying and discussing outlier cases; at a minimum, it should be standard practice to identify such cases.

Theory testing is perhaps the most common goal of combining formal and statistical methods. Good examples here include Downs, Rocke, and Barsoom (\hyperlink{ref-FMSBTT87}{1998}), Iversen (\hyperlink{ref-ZP9VBKJR}{1998}), Epstein and O'Halloran (\hyperlink{ref-9CS9L355}{1996}), and Lohmann and O'Halloran (\hyperlink{ref-EA23L7PE}{1994}). These examples underscore that a general challenge in statistical tests of formal models is that standard statistical models, and even more importantly standard statistical databases, do not typically allow seamless statistical testing of exactly the same model as specified in the formal work. The examples cited here generally devote great attention to this problem and go to considerable efforts to adapt standard databases to the variables defined in the formal model, often carrying out new codings of key variables or modifying existing measures. Without detailed knowledge of the databases in question it is impossible to judge how successful the authors have been or how they might have done better in testing the actual specifications of the formal model---our purpose here is merely to draw attention to this as the central challenge in this combination of methods.

Formal-statistical theory testing also appears to give only infrequent attention to fully specified alternative explanations. To some degree, this is understandable in view of the difficulty of specifying even a single formal model and testing it statistically, and in view of space constraints in journals. Also, at present most formal models have been developed in ``high density'' research programs, or programs that have attracted the attention of many researchers. As formal models become more common in a wider range of domains, however, more attention needs to be given to testing alternative explanations, whether formal or otherwise.

\subsection*{Combining Formal and Case Study Methods}

Case studies can evaluate both the predictions of formal models and their assumptions and internal logic. Game-theoretic models imply claims about actors' beliefs, possible strategies, and payoff structures to generate equilibrium predictions. Case studies can then be employed to assess whether these implied claims are empirically accurate in certain historical instances. Regardless of whether a formal model's predictions hold theoretically, it would be undermined if, for instance, it assumed that actors had perfect information only for case study evidence to reveal substantial uncertainty. This combination thus allows formal models to serve as generators of empirically testable propositions both for mechanisms as well as for outcomes. From this perspective, a formal model that correctly predicts an outcome under the wrong assumptions is as dangerous as one that predicts incorrectly. Case study investigation is well suited to distinguishing these possibilities (\hyperlink{ref-SRSGUSQK}{Lorentzen et al. 2017}).

Formal models can be used heuristically to derive deductively new models that can be applied to case studies. In much of this kind of research, the case studies are chosen to serve as illustrations or demonstrations of the formal model, without necessarily being full tests. In other words, the cases are often (but not always) chosen as motivating examples or most-likely cases, rather than being least-likely cases or tough tests of the formal model.

Conversely, case studies can be used inductively to develop new theories or variables that can be used to build formal models. This approach has been relatively neglected, however, in part due to the fact noted above that work using formal models frequently does not focus on outlier cases or even consider fully alternative explanations. The multi-authored book \emph{Analytic Narratives}, for example, while exemplary in many respects and laudable in its efforts to combine formal models and case studies, starts out with rational choice formal models and gives attention to alternative explanations only to explain the considerable variance that remains in each of the empirical chapters after rational choice models have been given their due (\hyperlink{ref-SZX5TU6K}{Bates et al. 1998}). In several chapters, important preferences beyond the original formal model emerge only toward the end of the study, rather than being treated from the outset as complementary or alternative explanations deserving of equal consideration. Such explanations might be included in the attendant formal models (admittedly a challenging task), rather than relegated to explaining the leftover variation. Seawright and Koivu (\hyperlink{ref-QJM8YRXH}{2026, 81--84}) see this tension between parsimony and detail as general to formal--case study work. On our account, the parsimony a model requires is itself a working assumption, which narrower subtypes of a phenomenon can relax.

Case studies can also be used to test whether the causal mechanisms developed in formal models actually applied in historical cases. By contrast, statistical tests of formal models remain inherently ambiguous for any individual case. This is where formal--case study has a comparative advantage, as case-confirming evidence is often difficult to recover from cross-case covariation (\hyperlink{ref-UA3ZA7YM}{Goertz 2017, 9--10, 26, 175}). Process tracing can evaluate not only whether a model's predictions hold but whether and how its causal logic operated in specific instances. This potential is often underachieved, however, due to problems of case selection. Here again, the \emph{Analytic Narratives} volume illustrates this problem, as it states that ``our cases selected us, rather than the other way around'' (\hyperlink{ref-SZX5TU6K}{Bates et al. 1998, 13}). It is important to note that these authors' goal was primarily disciplined-configurative studies, rather than full theory testing, and their case selection was more appropriate for their own research purposes, but formal--case study works that aspire to theory testing must give sufficient attention to alternative explanations and choose cases that are not subject to selection bias and that pose tough tests for the theory of interest.

One issue that does not consistently receive sufficient attention in formal--case study work is the problem of generalizing from case studies. There might be a temptation to assume that the deductive nature of formal models automatically assures them some generality, but of course this is not an unproblematic assumption. Generalization also requires carefully specifying the population to which the model is to be applied, or put in other terms, the contextual conditions in which it is relevant. Taking on tough test cases provides a presumption that a model will apply in easier test cases, but this presumption cannot be accepted on faith. The generalizability of a model is even less clear when the model is applied to most-likely cases or when the author is not clear in arguing how likely a case is for the model or how tough a test it provides. Goertz's (\hyperlink{ref-UA3ZA7YM}{2017, chap. 8}) ``medium-N paradigm'' serves as one example of how to design case study research that systematically explores scope conditions through targeted case selection. For all these reasons, following up a combination of a formal model and case studies with a statistical analysis of a population can provide greater confidence in the model's generalizability.

\subsection*{Combining Case Study and Statistical Methods}

A common way to integrate statistical analysis with case studies is to analyze statistically a data set on a specified population and then use the results of that analysis to help choose cases for case studies. Selecting cases for close study on the basis of information from a statistical analysis is not a one-size-fits-all affair. Different case selection criteria are necessary for different theory-building purposes. Fairfield and Charman (\hyperlink{ref-8288HE6Y}{2022, chap. 12}) instead reduce case selection to a single principle, which is to choose the cases expected to discriminate most sharply between rival hypotheses (551--52, 560). In practice, they recommend typical cases that fit an established statistical relationship for studying the mechanisms behind it (575--76). They recommend deviant cases for identifying omitted variables or refining scope conditions (576), and cases near the boundaries of a theory's scope for testing its generalization (552--53).

As Gerring and Seawright (\hyperlink{ref-GHM9DK79}{2006}) argue, depending on one's theory-testing and theory-building purposes, one might choose from a statistical analysis cases for detailed study that are typical, deviant, influential, extreme, diverse, most likely or least likely, most-similar or most-different, or representative of a particular pathway to the outcome of interest. We discuss each of these in turn.

Gerring and Seawright define a ``typical'' case primarily as a case with a low error term relative to a statistical model, but it is important to also take into consideration whether the values of a case's independent and dependent variables are close to and potentially representative of a cluster of other cases. This is not the same as being a case with average values on its variables, as Gerring and Seawright discuss, as it depends on the distribution of the population---in a population with many wealthy people, many poor people, and few people in between, a person of average wealth is not representative. Also, as Gerring and Seawright note, one cannot assume that a case has a low error term because it fits the model. The size of the error term can be misleading if the statistical model is misspecified (\hyperlink{ref-GHM9DK79}{Gerring and Seawright 2006, 96--97}; \hyperlink{ref-BD7XLSSF}{Rohlfing 2008, 1496--99}). Measurement error, one omitted variable with large effects, or many omitted variables with small effects can make a case appear to be typical of the processes underlying a population when it in fact reflects different causal processes.

Although claims and measures of typicality are not foolproof, studying a case with a low error term and high similarity to many other cases in a population can help assess potential spuriousness by testing whether the causal mechanisms hypothesized as possible explanations of a statistical distribution are in fact evident in a case that is by all known information representative of the population. In addition, the study of such cases, and the sequences through which their outcomes arose, can help address the potential for endogeneity by establishing whether the independent variables of interest appeared to precede and cause the dependent variable, or the reverse. Moreover, although it is not possible to eliminate the danger of model misspecification, case studies that proceed inductively as well as deductively, and those that attempt to trace backward from observed effects to possible causes as well as forward from hypothesized causes to observed effects, provide the potential for uncovering any omitted variables that might make a deviant case wrongly appear to be typical (\hyperlink{ref-BD7XLSSF}{Rohlfing 2008}).

In contrast to a typical case, a deviant case is one with a high error term relative to a statistical model; whether the processes behind such a case are or are not representative of those in the rest of the population is difficult to assess a priori, but this can be assessed by subsequent statistical analysis. Gerring and Seawright, among others, argue that the study of deviant cases provides an opportunity to look for explanations or variables left out of a statistical model (\hyperlink{ref-GHM9DK79}{Gerring and Seawright 2006, 106--7}).\footnote{Seawright's later work qualifies this advice. Cases that are extreme on the main causal variable are better than deviant cases for discovering important confounders (\hyperlink{ref-54ZCW7YV}{Seawright 2016, 86--101}; \hyperlink{ref-7YXV537F}{Galvin and Seawright 2023}; \hyperlink{ref-QJM8YRXH}{Seawright and Koivu 2026, 103}).}

Influential cases are a third kind of case that can be usefully selected from a statistical distribution for closer study. In contrast to analysis of a deviant case, the study of an influential case primarily seeks to affirm or disconfirm the validity of a statistical model, rather than to uncover omitted variables, although it can have the latter result as well (\hyperlink{ref-GHM9DK79}{Gerring and Seawright 2006, 109}). An influential case is one whose removal from a statistical sample would have a large effect on the regression estimates. For this reason, it is often important to study influential cases to assess whether their outcomes arose through the processes theorized as possible explanations of the correlations observed in a population. If closer study reveals that an influential case does not exhibit these processes, then the results of the statistical analysis are suspect and an argument can be made for removing the influential case from the statistical data set (presuming that there are strong theoretical and methodological reasons to believe that the hypothesized processes, and not those actually observed in the influential case, might explain the population).

The selection of diverse cases from a statistical distribution is designed to maximize variance on the variables of interest (\hyperlink{ref-GHM9DK79}{Gerring and Seawright 2006}; \hyperlink{ref-VCIMU9U8}{Lieberman 2005}). A related criterion for case selection is to select cases that have extreme values on the independent or dependent variables. An extreme value case may also have either a high or a low error term; the former constitutes a deviant as well as extreme case, and the latter may be an opportunity to observe in sharp relief causal mechanisms that may be relevant to the rest of the population (with the caveat noted above that a case with a small error is not necessarily typical or representative). The study of diverse cases is fairly common in multimethod works (\hyperlink{ref-6SXIFB58}{see, for example, Simmons 1994, 16--17}; \hyperlink{ref-WRUNG6NC}{Martin 1992, 99}). Choosing diverse cases can raise the likelihood that the results of the case study analysis will represent the population, or its stratified subsets, though there is a danger that the results will not be representative. For example, if there are many ``high'' cases and few ``low'' cases and the researcher chooses one of each, generalizing between the two can be misleading (\hyperlink{ref-GHM9DK79}{Gerring and Seawright 2006, 100}). There is a greater danger of inferential error in just choosing one extreme case and generalizing the results to a population (\hyperlink{ref-9LEVV6Q2}{Collier and Mahoney 1996}). Here, the argument for generalization can be strengthened by demonstrating that the processes behind the outcome of an extreme case were also evident in a typical case.

Many of these approaches to case selection can create misleading results if the researcher, in the absence of additional evidence, generalizes the case study findings to broad populations. This is why case study researchers often limit their findings to ``contingent generalizations,'' or generalizations on a well-defined subset of a population. At the same time, subsequent statistical research can incorporate the findings and variables of case studies, and additional case studies on typical or diverse cases can help establish the validity and scope conditions of case study findings. The combination of statistical and case study methods, in other words, can ease the necessity in statistical studies of making strong unit homogeneity assumptions or studying only subtypes for which sufficient examples exist for statistical methods.

\section*{Conclusion: Encouraging and Improving Multimethod Research}

Combining research methods can contribute greatly to progressive theorizing. Formal-statistical work can bring together the deductive theory-development power of formal models and the theory-testing power of statistical methods to derive and generalize counterintuitive insights. Formal--case study work can use both deductive and inductive methods to develop new theories and test them against tough cases. Case study--statistical work can provide convincing evidence that hypothesized causal mechanisms were operative in historical cases and that these mechanisms generalize to wider populations.

The increasingly evident complementarity of case studies, statistical methods, and formal models, and the continuing improvement of graduate training in each of these methods, provide the potential for increasingly sophisticated multimethod research. As each individual method becomes more sophisticated, however, it may become more difficult for any single researcher to be adept at more than one set of methods while also attaining a cutting-edge theoretical and empirical knowledge of their field. Successful collaboration is therefore likely to take the form of several researchers working together using different methods, as in many of the examples discussed above, or of researchers more self-consciously building on the findings generated by those using different methods. In either form, effective collaboration requires that even as they become expert in one methodological approach, scholars must also become conversant with alternative approaches, aware of their strengths and limits, and capable of an informed reading of their substantive results.

Unfortunately, for a variety of historical and institutional reasons, cross-method collaboration is not as well rewarded in the discipline as it should be. Graduate curricula do not consistently require or even offer courses in all three methods, and case study methods courses in particular are underrepresented relative to the proportion of published work that uses these methods (\hyperlink{ref-2BYY2P5T}{Bennett et al. 2003}). Journals in the same subfield---such as the \emph{Journal of Conflict Resolution} and \emph{International Security---}often focus only on one or two methods (formal models and statistics in the former, and case studies in the latter) and infrequently cite one another.

Further progress will require paying better attention to the evaluation of integrative research design. Reviewers primarily trained in specific traditions may end up holding multimethod work to standards that reflect working assumptions from their backgrounds. For example, a statistician could discredit a design that selects on the dependent variable even if the researcher chose deviant cases deliberately to search for omitted variables. Alternatively, a case study specialist may treat one or more mispredicted cases as disconfirming a model merely claiming to estimate average effects. Evaluating integrative research thus requires understanding how each component contributes to joint inference. Recent integrative multimethod frameworks have helped to make clearer what this process looks like in practice, but disciplinary norms continue to lag.

Increasing and improving multimethod work will require important changes in the discipline. The disincentives attached to multi-authored works in the academic culture of political science need to be addressed. Such work is treated in tenure and salary reviews as if the author of a work co-authored by two people, for example, had done less than half the work of a sole-authored article, whereas anyone who has ever co-authored a work feels with some justification that both authors have done more than half the work of a single article. Authors and journal editors might consider making it standard practice to state in the first footnote or the preface the division of labor among co-authors to help address this problem, but wider changes in the culture of the discipline are needed as well. Cross-method training also has to be improved, both in regular graduate curricula offerings and requirements and in dedicated training programs, such as the Inter-university Consortium for Political and Social Research (ICPSR) program centered at the University of Michigan or the Institute for Qualitative and Multi-Method Research (IQMR) at Syracuse University, that are designed to improve scholars' skills in one method or another. Finally, researchers from all three methodological approaches need to collaborate in the hard work of improving the definitions, measurements and operationalizations, and codings of databases. As our concepts and definitions continue to evolve, these databases require continual updating, and formal modelers and case study researchers need to contribute their expertise on concepts and cases to the enormous investment of effort that statistical researchers have made in developing data sets. The disciplinary and technical challenges to multimethod work are daunting, but the exemplary multimethod research that is already available demonstrates that the results will be well worth the effort.

\section*{References}\label{sec:references}
\begingroup
\singlespacing
\raggedright
\setlength{\parindent}{-0.5in}\setlength{\leftskip}{0.5in}\setlength{\parskip}{0.6\baselineskip}
\noindent\hspace{-0.5in}\hypertarget{ref-8R35HYVF}{}Ahmed, Amel, and Rudra Sil. 2012. ``When Multi-Method Research Subverts Methodological Pluralism---or, Why We Still Need Single-Method Research.'' \emph{Perspectives on Politics} 10 (4): 935--53. \url{https://doi.org/10.1017/S1537592712002836}.

\hypertarget{ref-9DWGRC2H}{}Ahram, Ariel I. 2013. ``Concepts and Measurement in Multimethod Research.'' \emph{Political Research Quarterly} 66 (2): 280--91. \url{https://doi.org/10.1177/1065912911427453}.

\hypertarget{ref-B5X6DKI6}{}Bäck, Hanna, and Patrick Dumont. 2007. ``Combining Large-n and Small-n Strategies: The Way Forward in Coalition Research.'' \emph{West European Politics} 30 (3): 467--501. \url{https://doi.org/10.1080/01402380701276295}.

\hypertarget{ref-SZX5TU6K}{}Bates, Robert H., Avner Greif, Margaret Levi, Jean-Laurent Rosenthal, and Barry R. Weingast. 1998. \emph{Analytic Narratives}. Princeton University Press.

\hypertarget{ref-BQK8DWJB}{}Beach, Derek. 2020. ``Multi-Method Research in the Social Sciences: A Review of Recent Frameworks and a Way Forward.'' \emph{Government and Opposition} 55 (1): 163--82. \url{https://doi.org/10.1017/gov.2018.53}.

\hypertarget{ref-NX8M7JSM}{}Beach, Derek, and Jonas Gejl Kaas. 2020. ``The Great Divides: Incommensurability, the Impossibility of Mixed-Methodology, and What to Do about It.'' \emph{International Studies Review} 22 (2): 214--35. \url{https://doi.org/10.1093/isr/viaa016}.

\hypertarget{ref-LU5M6EPK}{}Bennett, Andrew. 2015. ``Found in Translation: Combining Discourse Analysis with Computer Assisted Content Analysis.'' \emph{Millennium} 43 (3): 984--97. \url{https://doi.org/10.1177/0305829815581535}.

\hypertarget{ref-2BYY2P5T}{}Bennett, Andrew, Aharon Barth, and Kenneth R. Rutherford. 2003. ``Do We Preach What We Practice? A Survey of Methods in Political Science Journals and Curricula.'' \emph{PS: Political Science and Politics} 36 (3): 373--78.

\hypertarget{ref-8EDRRZ96}{}Bennett, Andrew, and Colin Elman. 2006. ``Complex Causal Relations and Case Study Methods: The Example of Path Dependence.'' \emph{Political Analysis} 14 (3): 250--67. \url{https://doi.org/10.1093/pan/mpj020}.

\hypertarget{ref-XPSIAPBR}{}Bennett, Andrew, Joseph Lepgold, and Danny Unger. 1994. ``Burden-Sharing in the Persian Gulf War.'' \emph{International Organization} 48 (1): 39--75. \url{https://doi.org/10.1017/S0020818300000813}.

\hypertarget{ref-8HEDZWYC}{}Blair, Graeme, Alexander Coppock, and Macartan Humphreys. 2023. \emph{Research Design in the Social Sciences: Declaration, Diagnosis, and Redesign}. Princeton University Press.

\hypertarget{ref-MEVHZHP6}{}Brambor, Thomas, William Roberts Clark, and Matt Golder. 2006. ``Understanding Interaction Models: Improving Empirical Analyses.'' \emph{Political Analysis} 14 (1): 63--82. \url{https://doi.org/10.1093/pan/mpi014}.

\hypertarget{ref-WQHW3M49}{}Braumoeller, Bear F. 2003. ``Causal Complexity and the Study of Politics.'' \emph{Political Analysis} 11 (3): 209--33. \url{https://doi.org/10.1093/pan/mpg012}.

\hypertarget{ref-2TZR38YI}{}Braumoeller, Bear F. 2008. ``Systemic Politics and the Origins of Great Power Conflict.'' \emph{American Political Science Review} 102 (1): 77--93. \url{https://doi.org/10.1017/S0003055408080088}.

\hypertarget{ref-KHSUM2S3}{}Cederman, Lars-Erik. 2001. ``Agent-Based Modeling in Political Science.'' \emph{The Political Methodologist} 10 (1): 16--22.

\hypertarget{ref-9U5Q86AS}{}Chernoff, Fred. 2007. ``Critical Realism, Scientific Realism, and International Relations Theory.'' \emph{Millennium} 35 (2): 399--407. \url{https://doi.org/10.1177/03058298070350021701}.

\hypertarget{ref-PQ5PGUWT}{}Clarke, Kevin A., and David M. Primo. 2012. \emph{A Model Discipline: Political Science and the Logic of Representations}. Oxford University Press.

\hypertarget{ref-LFPCNBTR}{}Collier, David, and Steven Levitsky. 1997. ``Democracy with Adjectives: Conceptual Innovation in Comparative Research.'' \emph{World Politics} 49 (3): 430--51. \url{https://doi.org/10.1353/wp.1997.0009}.

\hypertarget{ref-9LEVV6Q2}{}Collier, David, and James Mahoney. 1996. ``Insights and Pitfalls: Selection Bias in Qualitative Research.'' \emph{World Politics} 49 (1): 56--91. \url{https://doi.org/10.1353/wp.1996.0023}.

\hypertarget{ref-FMSBTT87}{}Downs, George W., David M. Rocke, and Peter N. Barsoom. 1998. ``Managing the Evolution of Multilateralism.'' \emph{International Organization} 52 (2): 397--419. \url{https://doi.org/10.1162/002081898753162866}.

\hypertarget{ref-Y4Q2EFLU}{}Eckstein, Harry. 1975. ``Case Study and Theory in Political Science.'' In \emph{Handbook of Political Science}, edited by Fred I. Greenstein and Nelson W. Polsby, vol. 7. Addison-Wesley Publishing Company.

\hypertarget{ref-T9W6J9NE}{}Elman, Colin. 2005. ``Explanatory Typologies in Qualitative Studies of International Politics.'' \emph{International Organization} 59 (2): 293--326. \url{https://doi.org/10.1017/S0020818305050101}.

\hypertarget{ref-9CS9L355}{}Epstein, David, and Sharyn O'Halloran. 1996. ``The Partisan Paradox and the U.S. Tariff, 1877--1934.'' \emph{International Organization} 50 (2): 301--24. \url{https://doi.org/10.1017/S0020818300028575}.

\hypertarget{ref-8288HE6Y}{}Fairfield, Tasha, and Andrew E. Charman. 2022. \emph{Social Inquiry and Bayesian Inference: Rethinking Qualitative Research}. Cambridge University Press.

\hypertarget{ref-I8J294LG}{}Findley, Michael G., Kyosuke Kikuta, and Michael Denly. 2021. ``External Validity.'' \emph{Annual Review of Political Science} 24 (May): 365--93. \url{https://doi.org/10.1146/annurev-polisci-041719-102556}.

\hypertarget{ref-7YXV537F}{}Galvin, Daniel J., and J. Seawright. 2023. ``Surprising Causes: Propensity-Adjusted Treatment Scores for Multimethod Case Selection.'' \emph{Sociological Methods \& Research} 52 (4): 1632--80. \url{https://doi.org/10.1177/00491241211004632}.

\hypertarget{ref-6BUL5HFZ}{}George, Alexander L., and Andrew Bennett. 2005. \emph{Case Studies and Theory Development in the Social Sciences}. BCSIA Studies in International Security. MIT Press.

\hypertarget{ref-GHM9DK79}{}Gerring, John, and J. Seawright. 2006. ``Techniques for Choosing Cases.'' In \emph{Case Study Research: Principles and Practices}, by John Gerring. Cambridge University Press. \url{https://doi.org/10.1017/CBO9780511803123.008}.

\hypertarget{ref-UA3ZA7YM}{}Goertz, Gary. 2017. \emph{Multimethod Research, Causal Mechanisms, and Case Studies: An Integrated Approach}. Princeton University Press.

\hypertarget{ref-SPJZYPYM}{}Goertz, Gary, and James Mahoney. 2012. \emph{A Tale of Two Cultures: Qualitative and Quantitative Research in the Social Sciences}. Princeton University Press.

\hypertarget{ref-GGQP5X7F}{}Grzymala-Busse, Anna. 2011. ``Time Will Tell? Temporality and the Analysis of Causal Mechanisms and Processes.'' \emph{Comparative Political Studies} 44 (9): 1267--97. \url{https://doi.org/10.1177/0010414010390653}.

\hypertarget{ref-CDC9T5NE}{}Hafner-Burton, Emilie M., and James Ron. 2009. ``Seeing Double: Human Rights Impact through Qualitative and Quantitative Eyes.'' \emph{World Politics} 61 (2): 360--401. \url{https://doi.org/10.1017/S0043887109000136}.

\hypertarget{ref-6YDIYBTN}{}Hall, Peter A. 2003. ``Aligning Ontology and Methodology in Comparative Politics.'' In \emph{Comparative Historical Analysis in the Social Sciences}, edited by James Mahoney and Dietrich Rueschemeyer. Cambridge University Press.

\hypertarget{ref-W87IX4B8}{}Holland, Paul W. 1986. ``Statistics and Causal Inference.'' \emph{Journal of the American Statistical Association} 81 (396): 945--60. \url{https://doi.org/10.1080/01621459.1986.10478354}.

\hypertarget{ref-6EJI8FVR}{}Humphreys, Macartan, and Alan M. Jacobs. 2015. ``Mixing Methods: A Bayesian Approach.'' \emph{American Political Science Review} 109 (4): 653--73. \url{https://doi.org/10.1017/S0003055415000453}.

\hypertarget{ref-TC45B3MR}{}Humphreys, Macartan, and Alan M. Jacobs. 2023. \emph{Integrating Inferences: Causal Models for Qualitative and Mixed-Method Research}. Cambridge University Press.

\hypertarget{ref-ZP9VBKJR}{}Iversen, Torben. 1998. ``Wage Bargaining, Central Bank Independence, and the Real Effects of Money.'' \emph{International Organization} 52 (3): 469--504. \url{https://doi.org/10.1162/002081898550635}.

\hypertarget{ref-UFX8R23S}{}King, Gary, Robert O. Keohane, and Sidney Verba. 1994. \emph{Designing Social Inquiry: Scientific Inference in Qualitative Research}. Princeton University Press.

\hypertarget{ref-ZRJSESUW}{}Kreuzer, Marcus. 2023. \emph{The Grammar of Time: A Toolbox for Comparative Historical Analysis}. Methods for Social Inquiry. Cambridge University Press.

\hypertarget{ref-VDZPQ7P8}{}Kuorikoski, Jaakko, and Caterina Marchionni. 2016. ``Evidential Diversity and the Triangulation of Phenomena.'' \emph{Philosophy of Science} 83 (2): 227--47. \url{https://doi.org/10.1086/684960}.

\hypertarget{ref-QIAY7TUF}{}Kuorikoski, Jaakko, and Caterina Marchionni. 2023. ``Evidential Variety and Mixed-Methods Research in Social Science.'' \emph{Philosophy of Science} 90 (5): 1449--58. \url{https://doi.org/10.1017/psa.2023.34}.

\hypertarget{ref-VCIMU9U8}{}Lieberman, Evan S. 2005. ``Nested Analysis as a Mixed-Method Strategy for Comparative Research.'' \emph{American Political Science Review} 99 (3): 435--52. \url{https://doi.org/10.1017/S0003055405051762}.

\hypertarget{ref-FUQJL7CL}{}Lijphart, Arend. 1971. ``Comparative Politics and the Comparative Method.'' \emph{American Political Science Review} 65 (3): 682--93. \url{https://doi.org/10.2307/1955513}.

\hypertarget{ref-EA23L7PE}{}Lohmann, Susanne, and Sharyn O'Halloran. 1994. ``Divided Government and U.S. Trade Policy: Theory and Evidence.'' \emph{International Organization} 48 (4): 595--632. \url{https://doi.org/10.1017/S0020818300028320}.

\hypertarget{ref-SRSGUSQK}{}Lorentzen, Peter, M. Taylor Fravel, and Jack Paine. 2017. ``Qualitative Investigation of Theoretical Models: The Value of Process Tracing.'' \emph{Journal of Theoretical Politics} 29 (3): 467--91. \url{https://doi.org/10.1177/0951629816664420}.

\hypertarget{ref-U6ZZ5C9U}{}Mahoney, James. 2000. ``Path Dependence in Historical Sociology.'' \emph{Theory and Society} 29 (4): 507--48.

\hypertarget{ref-T3FGLHWP}{}Mahoney, James, and Gary Goertz. 2006. ``A Tale of Two Cultures: Contrasting Quantitative and Qualitative Research.'' \emph{Political Analysis} 14 (3): 227--49. \url{https://doi.org/10.1093/pan/mpj017}.

\hypertarget{ref-WRUNG6NC}{}Martin, Lisa L. 1992. \emph{Coercive Cooperation: Explaining Multilateral Economic Sanctions}. Princeton University Press.

\hypertarget{ref-4LFCHVCD}{}Morgan, Stephen L., and Christopher Winship. 2014. \emph{Counterfactuals and Causal Inference: Methods and Principles for Social Research}. 2nd edition. Cambridge University Press.

\hypertarget{ref-55UHRDCB}{}Morton, Rebecca B. 1999. \emph{Methods and Models: A Guide to the Empirical Analysis of Formal Models in Political Science}. Cambridge University Press.

\hypertarget{ref-RWVRXHLP}{}Pierson, Paul. 2004. \emph{Politics in Time: History, Institutions, and Social Analysis}. Princeton University Press.

\hypertarget{ref-NDCA5MPS}{}Ragin, Charles C. 1987. \emph{The Comparative Method: Moving Beyond Qualitative and Quantitative Strategies}. University of California Press.

\hypertarget{ref-TQGN6566}{}Ragin, Charles C. 2000. \emph{Fuzzy-Set Social Science}. University of Chicago Press.

\hypertarget{ref-X6G2SIW5}{}Reiter, Dan, and Allan C. Stam. 2002. \emph{Democracies at War}. Princeton University Press.

\hypertarget{ref-BD7XLSSF}{}Rohlfing, Ingo. 2008. ``What You See and What You Get: Pitfalls and Principles of Nested Analysis in Comparative Research.'' \emph{Comparative Political Studies} 41 (11): 1492--514. \url{https://doi.org/10.1177/0010414007308019}.

\hypertarget{ref-VPMVSA8W}{}Rosenbaum, Paul R., and Donald B. Rubin. 1983. ``The Central Role of the Propensity Score in Observational Studies for Causal Effects.'' \emph{Biometrika} 70 (1): 41--55. \url{https://doi.org/10.1093/biomet/70.1.41}.

\hypertarget{ref-LRF4UJB6}{}Rubin, Donald B. 1973. ``Matching to Remove Bias in Observational Studies.'' \emph{Biometrics} 29 (1): 159--83. \url{https://doi.org/10.2307/2529684}.

\hypertarget{ref-8E9EIDTF}{}Rubin, Donald B. 1974. ``Estimating Causal Effects of Treatments in Randomized and Nonrandomized Studies.'' \emph{Journal of Educational Psychology} 66 (5): 688--701. \url{https://doi.org/10.1037/h0037350}.

\hypertarget{ref-JBAPDPC9}{}Sambanis, Nicholas. 2004. ``Using Case Studies to Expand Economic Models of Civil War.'' \emph{Perspectives on Politics} 2 (2): 259--79. \url{https://doi.org/10.1017/S1537592704040149}.

\hypertarget{ref-54ZCW7YV}{}Seawright, J. 2016. \emph{Multi-Method Social Science: Combining Qualitative and Quantitative Tools}. Cambridge University Press.

\hypertarget{ref-QJM8YRXH}{}Seawright, J., and Kendra L. Koivu. 2026. \emph{The Practice of Multi-Method Research: Mixing Qualitative and Quantitative Methods}. Cambridge University Press.

\hypertarget{ref-TMDYM4QZ}{}Signorino, Curtis S. 1999. ``Strategic Interaction and the Statistical Analysis of International Conflict.'' \emph{American Political Science Review} 93 (2): 279--97. \url{https://doi.org/10.2307/2585396}.

\hypertarget{ref-6SXIFB58}{}Simmons, Beth A. 1994. \emph{Who Adjusts?: Domestic Sources of Foreign Economic Policy During the Interwar Years}. Princeton University Press.

\hypertarget{ref-3Z7UTEAC}{}Small, Mario Luis. 2011. ``How to Conduct a Mixed Methods Study: Recent Trends in a Rapidly Growing Literature.'' \emph{Annual Review of Sociology} 37 (August): 57--86. \url{https://doi.org/10.1146/annurev.soc.012809.102657}.

\hypertarget{ref-Y7NHXXZ5}{}Smith, Alastair. 1999. ``Testing Theories of Strategic Choice: The Example of Crisis Escalation.'' \emph{American Journal of Political Science} 43 (4): 1254--83. \url{https://doi.org/10.2307/2991827}.

\hypertarget{ref-9I7Q33MH}{}Waldner, David. 2026. \emph{Qualitative Causal Inference and Explanation}. Cambridge University Press.

\hypertarget{ref-BMXZAIYA}{}Weibull, Jörgen W. 1997. \emph{Evolutionary Game Theory}. MIT Press.

\endgroup

\end{document}